\documentclass[twocolumn,prd,aps,showpacs,amsmath,amssymb,superscriptaddress,nofootinbib]{revtex4-1}
\usepackage{graphics}
\usepackage{graphicx}
\usepackage{dcolumn}
\usepackage{bm}
\usepackage{mathrsfs}
\usepackage{pstricks}
\usepackage{color}
\usepackage{slashed}
\usepackage{amsmath}
\usepackage{epsfig}
\usepackage{amsfonts}
\usepackage{amssymb}

\def\sgn{\mathrm{sgn}}
\def\beq{\begin{equation}}
\def\eeq{\end{equation}}
\def\bea{\begin{eqnarray}}
\def\eea{\end{eqnarray}}
\def\nn{\nonumber}

\def\Tr{\textrm{Tr}}
\def\x{\mathbf{x}}

\def\k{\mathbf{k}}

\begin{document}

\title{Entanglement dynamics in random media}
\author{G. Menezes}
\email{gsantosmenez@umass.edu}
\affiliation{Department of Physics, University of Massachusetts, Amherst, Massachusetts 01003, USA}
\affiliation{Departamento de F\'isica, Universidade Fe\-de\-ral Rural do Rio de Janeiro, 23897-000 Serop\'edica, RJ, Brazil}
\author{N. F. Svaiter}
\email{nfuxsvai@cbpf.br}
\affiliation{Centro Brasileiro de Pesquisas F\'{\i}sicas, Rio de Janeiro, RJ 22290-180, Brazil}
\author{C. A. D. Zarro}
\email{carlos.zarro@if.ufrj.br}
\affiliation{Instituto de F\'isica, Universidade Federal do Rio de Janeiro, Rio de Janeiro, RJ 21941-972, Brazil}

\begin{abstract}
We study how the entanglement dynamics between two-level atoms is impacted by random fluctuations of the light cone. In our model the two-atom system is envisaged as an open system coupled with an electromagnetic field in the vacuum state. We employ the quantum master equation in the Born-Markov approximation in order to describe the completely positive time evolution of the atomic system. We restrict our investigations to the situation in which the atoms are coupled individually to two spatially separated cavities, one of which displaying the emergence of light-cone fluctuations. In such a disordered cavity, we assume that the coefficients of the Klein-Gordon equation are random functions of the spatial coordinates. The disordered medium is modeled by a centered, stationary and Gaussian process. We demonstrate that disorder has the effect of slowing down the entanglement decay. We conjecture that in a strong disorder environment the mean life of entangled states can be enhanced in such a way as to almost completely suppress quantum nonlocal decoherence.
\end{abstract}


\maketitle

\section{Introduction}
\label{intro}

Entangled states are linear superpositions of quantum states that cannot be separated into product states of the individual systems. This property is long recognized as being a fully quantum-mechanical effect. Throughout the years entanglement has played a crucial role in many researches on the nature of quantum measurements, especially in the developments of quantum communications~\cite{1,haroche}.

Current investigations on atomic radiative processes have proved to be of paramount significance in quantum optics. Recent examples are connected with studies of radiative processes involving entanglement. For instance, in Ref.~\cite{yang} the authors investigate two entangled atoms coupled with an electromagnetic field in unbounded space. On the other hand, in Ref.~\cite{eberly} it was shown how vacuum fluctuations promote the entanglement decay of two qubits in a finite time. Ref.~\cite{yun} establishes a scheme to construct a highly controlled atom-field coupling in a cavity-quantum-electrodynamics system. Nonlinear optical processes in chains of ions in an entangled state were studied in Ref.~\cite{aga}. More recently, in a series of papers the radiative processes of entangled atoms were investigated in different physical settings~\cite{pra15,pra16,jhep,prd}. We observe that radiative processes of maximally entangled states should be distinguished from the non-entangled states~\cite{rep}.

All realizations of entangled atoms must confront with the decoherence phenomenon that is present in open systems. In other words, every operation of entanglement must furnish a strong coherent coupling between the atoms in order to minimize as much as possible the effects of decoherence coming from the interaction with the environment. Complications related with the endeavors of isolating the atoms from the environment emerge as the prime barrier preventing practical applications of entangled states. 

In this work we demonstrate a possible way to protect entanglement from a particular kind of decoherence, or at least to reduce the associated entanglement decay. Here we consider a system formed by two identical two-level atoms (which we call qubits) that are coupled individually with two spatially separated cavities. We suppose that in one of such cavities the structure of the light cones is altered due to the presence of disorder. Consequently, the eletromagnetic vacuum fluctuations in each cavity behave differently.

The interplay between disorder and quantum entanglement have been intensively studied in the recent literature. Let us briefly mention some of such important works. For instance, Anderson localization mechanism was reported in experiments involving entangled photons~\cite{disent1}. In a somewhat related investigation, the study of a disordered quantum Ising chain was explored in order to gather information on many-body localization transition~\cite{disent4}. In the context of quantum phase transitions, in Ref.~\cite{disent2} the entanglement entropy of random quantum critical points was investigated. The influence of disorder on quantum computing performance is studied in Ref.~\cite{disent5} in the context of spin-$1/2$ lattice models. With the objective of providing an efficient survey for experiments with quantum devices that could be appropriate for implementation of quantum computing, the authors in Ref.~\cite{disent3} analysed the entanglement dynamics of arrays of qubits in the presence of disorder. In all these papers the role of random fluctuations in quantum entanglement is uncovered in several different settings, each of which exploring a crucial aspect associated with disorder in the physical processes under consideration. 

This paper is devoted to provide further contributions to the aforementioned discussion by considering a framework which was not duly appreciated in the literature. Here we are interested in assessing the impact of light-cone fluctuations on the nonlocal decoherence of initially entangled atoms. Let us discuss what one means by ``light-cone fluctuations" in the present context. Ford and collaborators~\cite{ford1,ford2,ford3,ford4} have argued that fluctuations of the geometry of the space-time caused by a bath of squeezed gravitons has the effect of smearing out the light cone. On the other hand, light propagation in a nonlinear dielectric medium can be envisaged as a model of various effects involving quantum theory and gravity, such as the foregoing alluded light-cone fluctuations. In this case this involves the fluctuations of a background electric field, which in turn induce fluctuations in the ffective speed of light of probe pulses. This phenomenon was fully explored in Refs.~\cite{ford5,ford6}, in which the authors proposed the use of nonlinear dielectrics as analog models for light-cone fluctuations. On the other hand, in the lack of a precise knowledge of the nature of the fluctuations of a certain background field, one may assume that such fluctuations can be treated classically and their effects on the propagation of signals associated with quantum fields can be suitably modeled via random differential equations~\cite{pe13}. Such an alternative proposal was studied in a series of works~\cite{pe12,pe21,ellis,Bessa}. Within the scenario set up in such papers, one can address the issue of quantum entanglement in disordered media in a simple and straightforward way. Hence in this work the adoption of a wave equation with a random coefficient will be considered in order to investigate the influence of the disorder upon the entanglement dynamics of qubits. This will be discussed at length in due course. 

The aim of this work is to consider two initially entangled qubits and analyze the dynamics of their disentanglement when one of them is coupled with a disordered electromagnetic cavity. A possible realistic experimental realization of the present model is furnished by the experimental configurations advanced by Refs.~\cite{experimental1,experimental2}, where it is claimed a potential generation of a perfect entangled state between atoms in distant cavities. After allowing the interaction of one of such entangled atoms with a random dieletric medium, one obtains a feasible experimental contruction of the situation studied in this paper. We show that nonlocal decoherence caused by vacuum fluctuations of the quantum field is attenuated in the presence of random fluctuations of the light cone. The organization of this paper is as follows. In Sec.~\ref{lcf} we derive the positive-frequency Wightman function of the electric field in the presence of light-cone fluctuations. We work in the Coulomb gauge. We employ a perturbative expansion in order to take into account the effects of random fluctuations. In Sec.~\ref{master} we discuss the quantum master equation which describes the time evolution of the reduced density matrix of the two-qubit system. Sec.~\ref{entanglement} is devoted to the presentation and examination of the main results of this paper. Conclusions are given in Sec.~\ref{conclude}. We use units such that $\hbar=c=k_{B}=1$. Summation over repeated indices is assumed, unless otherwise stated. Graphics were drawn using the Mathematica packages.

\section{Light-cone fluctuations in electrodynamics}
\label{lcf}

As discussed in the previous section, we take the following situation. One of the atoms is immersed in a disordered medium in such a way that the associated photon propagation is described by a random wave equation. Being more specific, these photons evolve with random velocities. The other atom is inside an ``empty" cavity (in the sense of not having a noise field). We consider that both atoms follow inertial world lines. We are working in the interaction picture and in a four-dimensional Minkowski space-time. Let us consider the generic Hamiltonian of the two-atom system as:
\beq
H_A(\tau) = \frac{\omega_0}{2}\,{\bf n}\cdot\biggl[\boldsymbol\sigma_{(1)}(\tau) + \boldsymbol\sigma_{(2)}(\tau)\biggr],
\label{ha}
\eeq
where $\boldsymbol\sigma_{(1)} = \boldsymbol\sigma\otimes\sigma_0$, $\boldsymbol\sigma_{(2)} = \sigma_0\otimes \boldsymbol\sigma$, $\sigma^0$ is the unit $2 \times 2$ unit matrix, $\sigma^j$, $j = 1, 2, 3$ are the Pauli matrices and 
${\bf n} = (0,0,1)$ in Cartesian coordinates. Also, $\tau$ is the proper time of the atoms, while $\omega_0$ represents the gap between the two energy eigenvalues. 

The qubits are coupled with an electromagnetic field. In addition, we consider the case of atoms sufficiently far away from the walls of the cavities in such a way that their effects may be ignored. By exploiting the usual gauge invariance of the electromagnetic Lagrangian, we employ the Coulomb gauge. Using the usual decomposition of the electric and magnetic field operators in terms of creation and annihilation operators, the free electromagnetic field Hamiltonian can be written as
\beq
H_F(t) = \sum_{\lambda}\int \frac{d^3 k}{(2\pi)^3}\, \omega_{{\bf k}}\,a^{\dagger}_{{\bf k},\lambda}(t)a_{{\bf k},\lambda}(t),
\label{hf}
\eeq
where $\omega_{{\bf k}} = |{\bf k}|$ and $a^{\dagger}_{{\bf k},\lambda}, a_{{\bf k},\lambda}$ are the creation and annihilation operators of the electromagnetic field. We neglect the zero-point energy. The index $\lambda$ classifies different polarizations of the field. Orthonormal vectors ${\hat \varepsilon}_{\lambda}({\bf k})$, $\lambda = 1, 2$, correspond to the two polarization states of the photon. These obey known relations that can be found elsewhere~\cite{cohen}. We also take the electric field as divergenceless.

Let us discuss the interaction-picture Hamiltonian of interaction between the atoms and the electromagnetic field. Within the multipolar coupling scheme and in the dipole approximation one has that
\beq
H_{I}(\tau) = -{\bf P}_{(1)}(\tau)\cdot {\bf E}(x_1(\tau)) - {\bf P}_{(2)}(\tau)\cdot {\bf E}(x_2(\tau)),
\label{hi}
\eeq
where ${\bf P}_{(k)}(\tau)$ is the atomic dipole moment operator. It can be written as
\beq
{\bf P}_{(k)}(\tau) = \sum_{\eta = \pm} e^{i\eta\omega_0 \tau} {\cal M}_{(k)}(\eta \omega_0), 
\eeq
where ${\cal M}_{(k)}(-\omega_0) = {\bf p}_{(k)}\sigma_{-\,(k)}$ and ${\cal M}_{(k)}(\omega_0) = {\bf p}^{*}_{(k)}\sigma_{+\,(k)}$, ${\bf p}_{(k)}$ being the transition matrix element of the dipole operator of the atom $k$ in the usual basis formed by the eigenvectors of $\sigma^{3}_{(k)}$. In addition, $\sigma_{\pm\,(k)} = (1/2)(\sigma_{(k)}^{1} \pm i \sigma_{(k)}^{2})$.

Henceforth we consider that atom $2$ is immersed in a medium with light-cone fluctuations. We take as a disordered medium a heterogeneous, non-dissipative, non-magnetic dielectric medium whose dielectric constant $\epsilon({\bf x})/\epsilon_{0} = 1 + \chi({\bf x})$ is real and positive, and fluctuates randomly~\cite{ake}. Then a random wave equation for the vector potential follows from the Maxwell's equations. In other words, in the Coulomb gauge one has the following random wave equation in a four dimensional space-time:
\begin{equation}
\left[\bigl(1+\,\chi({\bf x})\bigr) \frac{\partial^{2}}{\partial\,t^{2}}
- \Delta\right] {\bf A}(t,{\bf x})=0,
\label{nami20}
\end{equation}
where as usual ${\bf E} = - \partial {\bf A}/\partial t$ and $\Delta$ is the three dimensional Laplacian. We have imposed that $c^2 = 1/(\epsilon_0\mu_0) = 1$. Note that in this medium $(\nabla \chi) \cdot {\bf E} = 0$. In addition, observe that $\chi({\bf x})$ works like a local refractive index, in that it perturbs locally the wave speed due to the replacement
$$
\frac{\partial^{2}}{\partial t^{2}}\rightarrow (1+\chi({\bf x}))
\frac{\partial^{2}}{\partial t^{2}}.
$$
For the random function $\chi({\bf x}\,)$ we will take a zero-mean Gaussian random
function:
\begin{equation}
\langle \chi({\bf x}\,) \rangle_{\chi} = 0, 
\label{nami21}
\end{equation}
with a white-noise correlation function given by
\begin{equation}
\langle \chi({\bf x}\,)\chi({\bf x}\,') \rangle_{\chi} = \sigma^2\,\delta^{(3)}({\bf x}-{\bf x}\,'), 
\label{nami22}
\end{equation}
where $\sigma^2$ gives the intensity of random fluctuations. The symbol $\langle\, ...\,\rangle_{\chi}$ denotes an average over
all possible realizations of this random variable. Note that we assume a time-independent random function for inertial observers. In addition, we impose that the fields inside and outside the random medium do not couple with each other.

Following Refs.~\cite{pe12,pe21}, in the weak-disorder limit the random wave equation~(\ref{nami20}) can be solved using a perturbative expansion in the noise function. We also suppose that the noises are Gaussian distributed (we use the notation $\chi(\x_i) = \chi_i $), that is
\begin{eqnarray}
\langle\chi_{i_1}  \cdots \chi_{i_{2n}} \rangle_\chi  &=&
\langle \chi_{i_1} \chi_{i_2} \rangle_\chi \langle \chi_{i_3} \chi_{i_4} \rangle_\chi
\langle \chi_{i_{2n-1}} \chi_{i_{2n}} \rangle_\chi \nonumber \\
&& + \, {\rm permutations} ,
\label{wick}
\end{eqnarray}
and correlations of an odd number of noises vanish. 

Fourier transforms are suitable to find 
solutions of Eq. (\ref{nami20}). The Fourier transform of
${\bf A}$ on the time variable $t$ is given by
\begin{equation}
{\bf A}(t,\x\,) = \int \frac{d\omega}{2\pi} \,
e^{-i \omega t}\, {\bf A}(\omega,\x),
\label{FTt-phi}
\end{equation}
and on the space variable $\x$:
\begin{equation}
{\bf A}(\cdot ,\x ) = \int \frac{d\k}{(2\pi)^3} \,
e^{i \k \cdot \x } \, {\bf A}(\cdot,\k).
\label{FTr-phi}
\end{equation}
Moreover, Fourier transforms of the stationary noise
functions are defined as:
\begin{equation}
\chi(\x ) = \int \frac{d\k}{(2\pi)^3} \,
e^{i \k\cdot\x } \,
\chi(\k ).
\label{FT-noise}
\end{equation}
One finds that the Fourier components of the field
${\bf A}(\omega,\k)$ satisfy the following algebraic equation 
\begin{equation}
\int d\k' \, \left[L_0(\k,\k') + L_1(\k,\k')\right] \, {\bf A}(\omega,\k') = 0,
\label{KG-mom}
\end{equation}
where $L_0$ is a non-stochastic matrix whose elements are written as:
\begin{eqnarray}
L_{0}(\k,\k') = \left(\omega^{2}
- \k^{2}\right) \delta(\k -\k'),
\label{L0-mom}
\end{eqnarray}
and $L_1(\k,\k')$ is a stochastic matrix, with elements:
\begin{eqnarray}
L_{1}(\k,\k') =  \chi(\k -\k') \, \omega^2.
\label{L1-mom}
\end{eqnarray}
In $\x$-space, $L_0$ and $L_1$ are:
\begin{eqnarray}
L_{0}(\x) &=& \omega^{2} + \Delta ,
\label{nami32} \\
L_{1}(\x) &=& \chi(\x)\,\omega^{2}.
\label{nami33}
\end{eqnarray}
They play a role as integral convolution operators. Notice
that while $L_{1}$ is diagonal, $L_{0}$ is non-diagonal in
$\x$-space; however, the opposite situation occurs in  $\k$-space. In
terms of these operators, the random wave equation reads, in matrix form:
\begin{equation}
(L_{0}+\,L_{1}) \, {\bf A}(\omega,\cdot) = 0 .
\label{nami31}
\end{equation}
From the above equation, the full (operator-valued) Green's function
$G$ is defined as
\begin{equation}
G=( L_{0} + L_{1} )^{-1}.
\label{green}
\end{equation}
Assuming that the noises are ``weak",
a perturbative expansion for~$G$ by means of a Dyson formula can be easily defined:
\begin{eqnarray}
G &=& G^{0} - G^{0}\, L_1 \,G^{0} + G^{0}\, L_1 \,G^{0}\, L_1 \,G^{0}
+ \, \cdots \nonumber \\
&=& G^{0} - G^{0} \, \Sigma \, G  ,
\label{green1}
\end{eqnarray}
with the self-energy $\Sigma$ given by
\begin{equation}
\Sigma = L_1 - L_1 G_0 L_1 + \cdots \, ,
\label{self}
\end{equation}
and $G^{0}=\,L_{0}^{-1}$ is the unperturbed (operator-valued) Green's
function. Finally, the Feynman Green's function can be written as
\bea
\hspace{-0.25cm}
G_{ab}(x,x') &=& \langle 0_{\chi} |T[A_{a}(x)A_{b}(x')]| 0_{\chi}\rangle  =  G^{0}_{ab}(x-x') 
\nn\\
&+&\, \sum_{n=1}^{\infty}\int dz_1 \,G^{0}_{ak}(x-z_1){\cal G}^{(n)}_{kb}(z_1,x'),
\label{a1}
\eea
where $|0_{\chi}\rangle$ is a modified vacuum state in the presence of disorder, $G^{0}_{ab}(x-x')$, $a, b = 1,2,3$, is the usual Feynman Green's function without random fluctuations and $T$ is the Dyson time-ordering symbol. The contributions from the light-cone fluctuations are represented by the following quantity
\begin{equation}
{\cal G}^{(n)}_{kb}(z_1,x') = (-1)^{n}\prod_{j=1}^n {\cal D}(z_j)
\int dz_{j+1} G^{0}_{m_{j}m_{j+1}}(z_j,z_{j+1})
\label{genericterm}
\end{equation}
with ${\cal D} (x)$ being the random differential operator
\beq
{\cal D} (x) ={\cal D}(t,{\bf r\,}) = - \chi({\bf r\,}) \frac{\partial^2}{\partial t^2}.
\label{L1-coord}
\eeq
In Eq.~(\ref{genericterm}), it is to be understood that $z_{n+1} = x'$, $m_{1} = k$, $m_{n+1} = b$ and that there is no integration in $z_{n+1}$. Details on similar derivations concerning a massless scalar field can be found in Ref.~\cite{pe21}. 

In the Coulomb gauge, the free photon propagator is given by
\bea
G^{0}_{ij}(x-x') &=& \langle 0 |T[A_{i}(x)A_{j}(x')]| 0\rangle 
\nn\\
&=&\, \int \frac{d^{4}k}{(2\pi)^4}\,\frac{i}{k^2 + i\epsilon}\left(\delta_{ij} - \frac{k_{i}k_{j}}
{|{\bf k}|^2}\right)\,e^{-ik\cdot(x-x')}
\nn\\
\eea
where $|0\rangle$ is the Minkowski vacuum state, $d^{4}k = dk^{0}d{\bf k}$ and $k^2 = (k^{0})^2 - |{\bf k}|^2$. Hence the zeroth order Feynman Green's function for the electric field reads:
\beq
D^{0}_{ij}(x,x') = \int \frac{d^{4}k}{(2\pi)^4}\,\frac{i\,(k^{0})^2}{k^2 + i\epsilon}\left(\delta_{ij} - \frac{k_{i}k_{j}}
{|{\bf k}|^2}\right)\,e^{-ik\cdot(x-x')}.
\eeq
In what follows we will be interested in the positive-frequency Wightman function for the electric field. The zeroth order contribution for such a function is given by~\cite{cohen}
\bea
\hspace{-0.07cm}
 D^{0,+}_{ij}(x,x') &=& 
-\frac{1}{\pi^2}\frac{1}{\left[(\Delta t - i\epsilon)^2 - |\Delta{\bf x}|^2\right]^3}
\nn\\
&\times&\Bigl\{2(\Delta{\bf x})_i(\Delta{\bf x})_j - \delta_{ij}\left[|\Delta{\bf x}|^2 + (\Delta t - i\epsilon)^2\right]\Bigr\},
\nn\\
\eea
where $\Delta t = t - t'$ and $\Delta {\bf x} = {\bf x} - {\bf x}'$. For the case of a static atom the unperturbed Wightman function of the electric field is given by
\beq
D^{0,+}_{ij}(x(\tau), x(\tau')) = \frac{ \delta_{ij}}{\pi^2(\tau-\tau' - i\epsilon)^4}.
\label{free}
\eeq
Coming back to Eq.~(\ref{a1}), after performing the averages over the noise function, the Feynman propagator associated with the photon field can be written in the form of a Dyson equation:
\begin{equation}
\langle G_{ab} \rangle_{\chi} =  G^{0}_{ab} - G^{0}_{am}\Sigma_{mk} \, \langle G_{kb} \rangle_{\chi},
\label{p32}
\end{equation}
with $\Sigma$ being the self-energy. The form of $\Sigma$ is obtained from noise averaging up to the second-order contribution in the random function $\chi$ and can be written generically as
\begin{equation}
\langle \, G_{ab}(x,x') \rangle_{\chi}= G^{0}_{ab}(x-x')
+ \langle\bar G^{1}_{ab}(x,x')\rangle_{\chi}.
\label{gg}
\end{equation}
For $\langle\bar G^{1}_{ab}(x,x')\rangle_{\chi}$ one obtains
\begin{equation}
\hspace{-0.04cm}
\langle\bar G^{1}_{ab}(x,x')\rangle_{\chi} = \int\frac{d{\bf k}}{(2\pi)^3}
\int\frac{dk^{0}}{2\pi} \, e^{-ik\cdot(x-x')} \, \langle\bar G^{1}_{ab}(k^{0},{\bf k})\rangle_{\chi}
\label{FT-G1}
\end{equation}
where
\begin{eqnarray}
\langle\bar G^{1}_{ab}(k^{0},{\bf k})\rangle_{\chi} &=& 
- \left[\frac{i}{k^2 + i\epsilon}\left(\delta_{aj} - \frac{k_{a}k_{j}}{|{\bf k}|^2}\right)\right]
\Sigma_{jm}(k^{0}) 
\nn\\
&\times&\,\left[\frac{i(k^{0})^2}{k^2 + i\epsilon}\left(\delta_{mb} - \frac{k_{m}k_{b}}{|{\bf k}|^2}\right)\right],
\label{eu}
\end{eqnarray}
with
\bea
\Sigma_{jm}(k^{0}) &=& - \sigma^2 \,(k^{0})^2\,\int \frac{d^{3}p}{(2\pi)^3}\,\frac{i}{(k^{0})^2 - |{\bf p}|^2 + i\epsilon}
\nn\\
&\times&\,\left(\delta_{jm} - \frac{p_{j}p_{m}}{|{\bf p}|^2}\right) 
\nn\\
&=&\, \sigma^2\,\frac{(k^{0})^3}{3\pi}\,\delta_{jm}.
\label{Sigma}
\eea
After inserting the above results in Eq.~(\ref{FT-G1}) one obtains:
\bea
\langle\bar G^{1}_{ab}(x,x')\rangle_{\chi} &=& \frac{\sigma^2}{3\pi}\,
\int\frac{d{\bf k}}{(2\pi)^3}\int\frac{dk^{0}}{2\pi} \, e^{-ik\cdot(x-x')} \, 
\nn\\
&\times&\,\frac{(k^{0})^5}{(k^2 + i\epsilon)^2}\left(\delta_{ab} - \frac{k_{a}k_{b}}{|{\bf k}|^2}\right).
\eea
The integral over $k^{0}$ can be evaluated within contour-integration methods. On the other hand, in order to present the corrections to the positive-frequency Wightman function, we only need to consider $t > t'$ and the pole at 
$k^{0} = |{\bf k}| - i\epsilon$. The contour we use is the usual contour for the positive-frequency Wightman function, see page $22$ of Ref.~\cite{birrel}. One gets
\bea
\hspace{-0.3cm}
\langle\bar G^{1,+}_{ab}(x,x')\rangle_{\chi} &=& \frac{\sigma^2}{12\pi}\,
\int\frac{d{\bf k}}{(2\pi)^3}|{\bf k}|^2  \Bigl(4 - i |{\bf k}|(t-t')\Bigr)
\nn\\
&\times&\left(\delta_{ab} - \frac{k_{a}k_{b}}{|{\bf k}|^2}\right)
e^{i{\bf k} \cdot ({\bf x} - {\bf x}') - i|{\bf k}|(t-t')}. 
\label{wigh-A}
\eea
Hence the contribution to the positive-frequency Wightman function of the electric field is given by
\bea
\hspace{-0.25cm}
\langle\bar D^{1,+}_{ab}(x,x')\rangle_{\chi} &=& \frac{\sigma^2}{12\pi}\,
\int\frac{d{\bf k}}{(2\pi)^3}|{\bf k}|^4  \Bigl(6 - i |{\bf k}|(t-t')\Bigr)
\nn\\
&\times& \left(\delta_{ab} - \frac{k_{a}k_{b}}{|{\bf k}|^2}\right)
e^{i{\bf k} \cdot ({\bf x} - {\bf x}') - i|{\bf k}|(t-t')}. 
\label{wigh-E}
\eea
Evaluating the integrals one finds that
\begin{widetext}
\bea
\langle\bar D^{1,+}_{ab}(x,x')\rangle_{\chi} &=& \frac{4 i \sigma^2}
{\pi^3 \left[(\Delta t-i \epsilon )^2-|\Delta {\bf x}|^2\right]^7}
\left\{
12(\Delta{\bf x})_a(\Delta{\bf x})_b
\Bigl[2(\Delta t-i \epsilon)\left(3 |\Delta {\bf x}|^4 + 4 |\Delta {\bf x}|^2 (\Delta t-i \epsilon )^2 -7 (\Delta t - i \epsilon )^4\right)
\right.
\nn\\
&+&\, \left. \Delta t\left(|\Delta {\bf x}|^4 + 18 |\Delta {\bf x}|^2 (\Delta t - i \epsilon )^2 
+ 21 (\Delta t - i \epsilon )^4\right)\Bigr] \right.
\nn\\
&-&\, \left. \delta_{ab}
\Bigl[6(\Delta t-i \epsilon)\left(9 |\Delta {\bf x}|^6 + 17 |\Delta {\bf x}|^4 (\Delta t - i \epsilon )^2
-21 |\Delta {\bf x}|^2 (\Delta t - i \epsilon )^4 - 5 (\Delta t -i \epsilon )^6\right)
\right.
\nn\\
&+&\, \left. \Delta t\left(9|\Delta {\bf x}|^6 + 177 |\Delta {\bf x}|^4(\Delta t - i \epsilon )^2
+259  |\Delta {\bf x}|^2 (\Delta t-i \epsilon )^4 + 35 (\Delta t -i \epsilon )^6\right)\Bigr]\right\}.
\eea
\end{widetext}
For a static atom, one has that
\beq
\left\langle\bar D^{1,+}_{ij}(x(\tau), x(\tau'))\right\rangle_{\chi} = -\frac{20\, i \sigma^2\,\delta_{ij}}
{\pi^3 (\tau - \tau' - i \epsilon )^{7}}.
\label{disorder}
\eeq
%

\section{The master equation approach}
\label{master}

In this section we discuss the time evolution of the qubit system coupled with a quantized electromagnetic field within the master equation approach. A similar analysis is encountered in Ref.~\cite{yu}, in which the authors study entanglement dynamics of two uniformly accelerated atoms coupled with an electromagnetic field within master-equation techniques. We will consider the so-called Born approximation in which initial interaction-induced correlations between the subsystems are neglected. This is justified as long as the spectral width of the quantum field energies is much broader than the spectral width associated with the interaction between the subsystems. We also consider the relaxation time associated with the noise field to be much larger than the reciprocals of the spectral widths aforementioned. This requirement is necessary in order to qualify the disorder as being of the quenched type.

Taking into account the Born approximation aforecited, one can take an initial density operator of the whole system of qubits and fields in a factorized form: $\rho(0) = \rho_{A}(0) \otimes \rho_{F}$, where $\rho_{A}(0)$ is the initial density operator associated with the qubit system whereas $\rho_{F}$ is the analogous one for the quantum electromagnetic field (assumed
to be stationary). The latter is given by $\rho_{F} = |\Omega\rangle\langle \Omega|$, where $|\Omega\rangle = |0\rangle$ for the empty cavity and $|\Omega\rangle = |0_{\chi}\rangle$ for the disordered cavity. On the other hand, for the initial state $\rho_{A}(0)$ we will consider an entangled state whose specific form will be expounded in due course. 

The subdynamics describing the evolution of the qubits subsystem alone is obtained with a trace operation over the field degrees of freedom. The details are discussed in many works~\cite{edavies1,edavies2,lindblad,koss,cohen4,rep,ben1,ben2,ben3}. In the limit of weak coupling between the atoms and the field, and within a suitable Markovian approximation, the evolution of the reduced density matrix $\rho_{A}(\tau)$ can be written in the Kossakowski-Lindblad form
\beq
\frac{\partial \rho_{A}(\tau)}{\partial \tau} = -i[H_{\textrm{eff}}, \rho_{A}(\tau)] + {\cal L}[\rho_{A}(\tau)],
\label{eq-master}
\eeq
where
\bea
{\cal L}[\rho_{A}(\tau)] &=& \frac{1}{2}\sum_{a,b = 1}^{2}\,{\cal K}_{ij}^{(a)(b)}\,\left\{2\sigma_{(b)}^{j}\,
\rho_{A}(\tau)\sigma_{(a)}^{i} \right.
\nn\\
&-&\, \left. \{\sigma_{(a)}^{i}\,\sigma_{(b)}^{j},\rho_{A}(\tau)\}\right\},
\label{nonunitary}
\eea
and the effective Hamiltonian reads
\beq
H_{\textrm{eff}} = H_A - \frac{i}{2}\sum_{a,b=1}^{2}{\cal H}_{ij}^{(a)(b)}\,\sigma_{(a)}^{i}\sigma_{(b)}^{j},
\label{eff}
\eeq
with 
\beq
{\cal H}_{ij}^{(a)(b)}(\omega_0) = d_{+}^{(a)(b)}\delta_{ij} - d_{+}^{(a)(b)}\,n_{i}n_{j} - i d_{-}^{(a)(b)}\,
\epsilon_{ijm}\,n^{m}.
\eeq
In the above equation $\epsilon_{ijk}$ is the usual Levi-Civita symbol. The Kossakowski matrix is given by
\beq
{\cal K}_{ij}^{(a)(b)}(\omega_0) = g_{+}^{(a)(b)}\delta_{ij} - g_{+}^{(a)(b)}\,n_{i}n_{j} - i g_{-}^{(a)(b)}\,
\epsilon_{ijm}\,n^{m}.
\label{koss}
\eeq
In the above expressions we have defined the quantities
\bea
g_{\pm}^{(a)(b)} &=& \frac{1}{4}\Bigl[{\cal G}^{(a)(b)}(\omega_0) \pm {\cal G}^{(a)(b)}(-\omega_0)\Bigr]
\nn\\
d_{\pm}^{(a)(b)} &=& \frac{1}{4}\Bigl[{\cal D}^{(a)(b)}(\omega_0) \pm {\cal D}^{(a)(b)}(-\omega_0)\Bigr]
\eea
where (no summation over $a,b$)
\bea
{\cal G}^{(a)(b)}(\omega_0) &=&  p^{i}_{(a)}p^{j\,*}_{(b)}{\cal G}^{(a)(b)}_{ij}(\omega_0)
\nn\\
{\cal D}^{(a)(b)}(\omega_0) &=& p^{i}_{(a)}p^{j\,*}_{(b)}{\cal D}^{(a)(b)}_{ij}(\omega_0)
\eea
In the above equations we have introduced the following Fourier and Hilbert transform of the electromagnetic field correlations
\beq
{\cal G}^{(a)(b)}_{ij}(\eta\omega_0) = \int_{-\infty}^{\infty}du\,e^{i\eta\omega_0 u}\,D^{(a)(b)}_{ij}(u),
\label{fourier}
\eeq
and 
\bea
{\cal D}^{(a)(b)}_{ij}(\eta\omega_0) &=&  \int_{-\infty}^{\infty}du\,e^{i\eta\omega_0 u}\,\sgn(u)\,D^{(a)(b)}_{ij}(u)
\nn\\
&=& \frac{\textrm{P}}{\pi i}\int_{-\infty}^{\infty}d\lambda\,\frac{{\cal G}^{(a)(b)}_{ij}(\lambda)}{\lambda - \eta\omega_0},
\label{hilbert}
\eea
where $\textrm{P}$ denotes the principal value and the correlation function 
\bea
D^{(a)(b)}_{ij}(\tau-\tau') &=& D_{ij}(x_a(\tau)-y_b(\tau'))
\nn\\
&=&\, \langle \Omega|E_{i}(x_a(\tau))E_{j}(y_b(\tau'))|\Omega\rangle.
\nn
\eea
In the above equations $\eta = \pm$. A suitable divergent (frequency-independent) term may need to be subtracted by a suitable renormalization procedure in order to make the contribution $H_{\textrm{eff}}$ well defined. This will be discussed below. A similar situation also appears in~\cite{ben2}. On the other hand, we recall that in the present situation the atoms are inside cavities whose fields do not interact with each other. This means that crossed terms in the Kossakowski matrix are absent in our analysis. 

Let us evaluate the quantities ${\cal G}^{(a)(b)}(\omega_0)$ and ${\cal D}^{(a)(b)}(\omega_0)$ which enter in the above definitions of the effective Hamiltonian and the Kossakowski matrix, respectively Eqs.~(\ref{eff}) and~(\ref{koss}). For the qubit inside the empty cavity, the positive-frequency Wightman function is given by Eq.~(\ref{free}). Inserting such an expression in Eq.~(\ref{fourier}) one obtains
\beq
{\cal G}^{(1)(1)}_{kl}(\eta\omega_0) = \frac{\eta ^3 \omega _0^3}{3 \pi}\,\theta(\eta)\,\delta_{kl}.
\eeq
On the other hand, concerning the other qubit which is coupled with the disordered cavity, the Fourier transform of the positive-frequency Wightman function of the electric field can be written as
\beq
{\cal G}^{(2)(2)}_{kl}(\eta\omega_0) = {\cal G}^{(2)(2)\,0}_{kl}(\eta\omega_0) 
+ \langle \bar {\cal G}^{(2)(2)\,1}_{kl}(\eta\omega_0) \rangle
\eeq
where ${\cal G}^{(2)(2)\,0}_{kl}(\eta\omega_0) = {\cal G}^{(1)(1)}_{kl}(\eta\omega_0)$ and
\beq
\langle \bar {\cal G}^{(2)(2)\,1}_{kl}(\eta\omega_0) \rangle = -\frac{\sigma ^2\eta^6\omega_0^6}{18 \pi ^2}\,\theta(\eta)\,\delta_{kl}.
\eeq
As discussed above one has ${\cal G}^{(a)(b)}_{kl}(0) = 0$ for $a \neq b$. In this way the explicit expression for the Kossakowski matrix reads
\beq
{\cal K}_{ij}^{(a)(b)}(\omega_0) = {\cal A}^{(a)(b)}[\delta_{ij} - n_{i}n_{j} - i  \epsilon_{ijm}\,n^{m}],
\label{koss1}
\eeq
where
\beq
{\cal A}^{(1)(1)} = \frac{\omega _0^3}{12 \pi}\Bigl(\delta_{ij}\,p^{i}_{(1)}p^{j\,*}_{(1)}\Bigr)
\label{a11}
\eeq
and
\beq
{\cal A}^{(2)(2)} = \frac{\omega _0^3}{12 \pi}\left(1 - \frac{\sigma ^2\omega_0^3}{6 \pi}\right)
\Bigl(\delta_{ij}\,p^{i}_{(2)}p^{j\,*}_{(2)}\Bigr).
\label{a22}
\eeq
with ${\cal K}_{ij}^{(a)(b)} = 0$ for $a \neq b$. Note that depending on the values of $\omega_0$ and $\sigma$, the Kossakowski matrix can acquire a negative eigenvalue, which means that an arbitrary initial state could be mapped out of the state-space by the time evolution~\cite{ben3}. The consequences of this adversity will be discussed below.

Concerning the Hilbert transform, one gets the following result for the empty cavity
\beq
{\cal D}_{kl}^{(1)(1)}(\eta\omega_0) = \frac{\delta_{kl}}{3 \pi}\frac{\textrm{P}}{\pi i}\int_{0}^{\infty}d\lambda\,\frac{\lambda^3}{\lambda - \eta\omega_0}.
\vspace{1mm}
\eeq
For the qubit inside the disordered cavity one finds that
\beq
{\cal D}_{kl}^{(2)(2)}(\eta\omega_0) = {\cal D}_{kl}^{(2)(2)\,0}(\eta\omega) 
+ \langle \bar {\cal D}_{kl}^{(2)(2)\,1}(\eta\omega_0) \rangle,
\vspace{1mm}
\eeq
with ${\cal D}_{kl}^{(2)(2)\,0}(\eta\omega_0) = {\cal D}_{kl}^{(1)(1)}(\eta\omega_0)$ and
\beq
\langle \bar {\cal D}_{kl}^{(2)(2)\,1}(\eta\omega_0) \rangle = -\frac{\delta_{kl}\,\sigma^2}{18 \pi^2}
\frac{\textrm{P}}{\pi i}\int_{0}^{\infty}d\lambda\,\frac{\lambda^6}{\lambda - \eta\omega_0}.
\eeq
Hence using that $\sigma^{(a)}_{i}\sigma^{(a)}_{j} = \delta_{ij}\,{\bf 1}^{(a)} +  i \epsilon_{ijk}\sigma^{(a)}_{k}$, together with $\epsilon_{aij}\,\epsilon^{bij} = 2 \delta_{a}^{\,\,b}$, the effective Hamiltonian can be written as
\beq
H_{\textrm{eff}} = \frac{\,\,\,\Omega^{(1)}}{2}\,{\bf n}\cdot\boldsymbol\sigma^{(1)} 
+ \frac{\,\,\,\Omega^{(2)}}{2}\,{\bf n}\cdot\boldsymbol\sigma^{(2)},
\eeq
where we have defined the renormalized frequencies $\Omega^{(a)}$, $a = 1, 2$, as (no summation over $a$)
\beq
\Omega^{(a)} = \omega_0 - \frac{i}{2}\,\Bigl[{\cal D}^{(a)(a)}(\omega_0) - {\cal D}^{(a)(a)}(-\omega_0)\Bigr].
\eeq
Note that the contributions ${\cal D}^{(a)(a)}(\eta\omega_0)$ produce the usual Lamb shift in the energy eigenvalues. As can be easily checked, the Lamb contribution turns out to be infinite, and its definition calls for the introduction of an adequate cutoff and a subsequent renormalization procedure. This is a well-known fact; the appearance of the divergences is related to the nonrelativistic treatment of the two-level atoms. It is mandatory the use of quantum-field-theory techniques for any sensible assessment of the energy shifts.

\section{Quantum entanglement in the presence of light-cone fluctuations}
\label{entanglement}

Now let us finally move on to the analysis of the contribution of light-cone fluctuations upon the quantum entanglement of inertial qubits in Minkowski spacetime. We find convenient to write the density matrix in the basis formed by the Dirac matrices~\cite{gamel}
\beq
\Sigma_{\mu\nu} = \sigma_{\mu}\otimes\sigma_{\nu}.
\eeq
In this way the reduced density matrix associated with the qubits subsystem can be written in a generic way as
\beq
\rho_{A}(\tau) = \frac{1}{4}\,\rho_{\mu\nu}(\tau)\Sigma_{\mu\nu},
\eeq
where $\rho_{\mu\nu}(\tau)$ is the Bloch matrix. With the normalization condition $\Tr[\rho_{A}(\tau)] = 1$, and imposing that $\rho_{A}(\tau)$ should be Hermitian, one has that $\rho_{00} = 1$ and $\rho_{i0}(\tau), \rho_{0i}(\tau), \rho_{ij}(\tau)$ are real. Explicitly ($i,j = 1,2,3$):
\bea
\rho_{A}(\tau) &=& \frac{1}{4}\left[\sigma^0\otimes\sigma^0 + \rho_{0i}(\tau)\,\sigma^{0}\otimes\sigma^{i}
+ \rho_{i0}(\tau)\,\sigma^{i}\otimes\sigma^{0} \right.
\nn\\
&+&\, \left. \rho_{ij}(\tau)\,\sigma^{i}\otimes\sigma^{j}\right]
\nn\\
&=&\,\frac{1}{4}\left[{\bf 1}_{4 \times 4} + \rho_{0i}(\tau)\,\sigma_{(2)}^{i}
+ \rho_{i0}(\tau)\,\sigma_{(1)}^{i} \right.
\nn\\
&+&\, \left. \rho_{ij}(\tau)\,\sigma_{(1)}^{i}\sigma_{(2)}^{j}\right].
\eea
In what follows we will ignore the Hamiltonian piece since it cannot give rise to entanglement phenomena, and focus attention on the study of the effects induced by the dissipative part ${\cal L}[\rho_{A}(\tau)]$. For the particular case in which ${\bf n} = (0,0,1)$, one gets the following evolution equations for the components of $\rho_{A}(\tau)$:
\bea
\frac{\partial\rho_{0k}}{\partial \tau} &=& - 2{\cal A}^{(2)(2)}\rho_{0k}(\tau),\,\,\, k \neq 3 
\nn\\
\frac{\partial\rho_{03}}{\partial \tau} &=& -4{\cal A}^{(2)(2)}\rho_{03}(\tau)  - 4 {\cal A}^{(2)(2)}
\nn\\
\frac{\partial\rho_{k0}}{\partial \tau} &=& -2{\cal A}^{(1)(1)}\rho_{k0}(\tau),\,\,\, k \neq 3  
\nn\\
\frac{\partial\rho_{30}}{\partial \tau} &=& -4{\cal A}^{(1)(1)}\rho_{30}(\tau) - 4 {\cal A}^{(1)(1)}
\eea
and finally
\begin{widetext}
\bea
\frac{\partial\rho_{ij}}{\partial \tau} &=& - 2({\cal A}^{(1)(1)} + {\cal A}^{(2)(2)})\rho_{ij}(\tau),\,\,\, i,j \neq 3
\nn\\
\frac{\partial\rho_{i3}}{\partial \tau} &=& - 2({\cal A}^{(1)(1)} + 2{\cal A}^{(2)(2)})\rho_{i3}(\tau)
 - 4 {\cal A}^{(2)(2)} \rho_{i0},\,\,\,\, i =1,2
\nn\\
\frac{\partial\rho_{3j}}{\partial \tau} &=& - 2(2{\cal A}^{(1)(1)} + {\cal A}^{(2)(2)})\rho_{3j}(\tau)
- 4 {\cal A}^{(1)(1)}\rho_{0j},\,\,\, j =1,2
\nn\\
\frac{\partial\rho_{33}}{\partial \tau} &=& - 4({\cal A}^{(1)(1)} + {\cal A}^{(2)(2)})\rho_{33}(\tau)
- 4 {\cal A}^{(1)(1)}\rho_{03}(\tau) - 4 {\cal A}^{(2)(2)}\rho_{30}(\tau).
\eea
The solutions of such differential equations are given by
\bea
\rho_{0i}(\tau) &=& \rho_{0i}(0)e^{-2{\cal A}^{(2)(2)}\tau},\,\,\, i = 1,2
\nn\\
\rho_{03}(\tau) &=& -1+ \Bigl[\rho_{03}(0) + 1 \Bigr]e^{-4{\cal A}^{(2)(2)}\tau}
\nn\\
\rho_{i0}(\tau) &=& \rho_{i0}(0)e^{-2{\cal A}^{(1)(1)}\tau},\,\,\, i = 1,2
\nn\\
\rho_{30}(\tau) &=& - 1 + \Bigl[\rho_{30}(0) + 1 \Bigr]e^{-4{\cal A}^{(1)(1)}\tau}
\nn\\
\rho_{ij}(\tau) &=& \rho_{ij}(0)e^{- 2({\cal A}^{(1)(1)} + {\cal A}^{(2)(2)})\tau},\,\,\, i,j \neq 3
\nn\\
\rho_{i3}(\tau) &=& \left[\rho_{i3}(0) + \rho_{i0}(0)\left(1- e^{4 {\cal A}^{(2)(2)} \tau}\right)\right] 
e^{-2 \tau ({\cal A}^{(1)(1)} + 2 {\cal A}^{(2)(2)})},\,\,\,\, i =1,2
\nn\\
\rho_{3j}(\tau) &=& \left[\rho_{3j}(0) + \rho_{0j}(0)\left(1 -  e^{4 {\cal A}^{(1)(1)} \tau}\right)\right] 
e^{-2 \tau (2{\cal A}^{(1)(1)} + {\cal A}^{(2)(2)})},\,\,\, j =1,2
\nn\\
\rho_{33}(\tau) &=& e^{-4 ({\cal A}^{(1)(1)} + {\cal A}^{(2)(2)})\tau} \left\{1 + \rho_{03}(0) + \rho_{30}(0) + \rho_{33}(0) +
e^{4 ({\cal A}^{(1)(1)} + {\cal A}^{(2)(2)})\tau} \right.
\nn\\
&-&\, \left. \Bigl[\rho_{03}(0) + 1 \Bigr] e^{4 {\cal A}^{(1)(1)} \tau} -  \Bigl[\rho_{30}(0) + 1 \Bigr] e^{4 {\cal A}^{(2)(2)} \tau}\right\}.
\eea
\end{widetext}
We consider the situation in which the atoms are initially prepared in an entangled state. Being more specific, let us consider the case in which the initial reduced density operator $\rho_{A}(0)$ is given by the $2$-qubit family of Werner states
\beq
\rho_{A}(0) = \kappa\,\rho^{-} + \left(\frac{1-\kappa}{4}\right)\sigma^0\otimes\sigma^0,
\label{initial-rho}
\eeq
where $\kappa$ is a real number and $\rho^{-}$ is the maximally entangled singlet state:
\beq
\rho^{-} = \frac{1}{4}\left[\sigma^0\otimes\sigma^0 - \delta_{ij}\,\sigma^{i}\otimes\sigma^{j}\right].
\eeq
For $\kappa > 1/3$ such a state is entangled. Notice that $\kappa$ is related with the trace $\rho_{ii}(0)$, $i=1,2,3$, of the initial density matrix given by Eq.~(\ref{initial-rho}).

In order to discuss the measure of the entanglement amount of the reduced density matrix $\rho_{A}(\tau)$, we resort to the concurrence ${\cal C}[\rho]$~\cite{conc1,conc2,conc3}, which is a monotonically increasing function of the entanglement of formation~\cite{conc4}. The concurrence takes values in the interval $[0,1]$, where zero is for separable states and one is for maximally entangled states, as the Bell states. In order to evaluate the concurrence of any $4\times4$ density matrix $\rho$ portraying the state of two qubits, one considers the square roots $\lambda$ of each of the four eigenvalues of the matrix $\rho(\sigma^{2}\otimes\sigma^{2})\rho^{{\cal T}}(\sigma^{2}\otimes\sigma^{2})$, where ${\cal T}$ denotes transposition. Then one considers such values in decreasing order: $\lambda_{1} \geq \lambda_{2} \geq \lambda_{3} \geq \lambda_{4}$. The concurrence of $\rho$ is defined to be ${\cal C}[\rho] = \textrm{max}\{\lambda_{1} - \lambda_{2} - \lambda_{3} - \lambda_{4},0\}$. 

In order to ensure that $\rho_{A}(\tau)$ has only positive eigenvalues for all finite $\tau$, one has to impose certain conditions over the possible values of $\kappa$ for fixed values of $\omega_0$, $\sigma^2$ and polarizations of the atoms. The simplest situation is to assume that $ \kappa \leq 1$. As a consequence, the concurrence will be a purely real number, as revealed by direct inspection of the square roots $\lambda$. In turn, the presence of imaginary values for the concurrence can also be seen as a consequence of the possible lack of positive-definiteness of the Kossakowski matrix discussed above. 

For $\kappa = 1$ ($\rho_{A}(0) = \rho^{-}$), the concurrence is given by the simple expression
\beq
{\cal C}[\rho_{A}(\tau)] = e^{-2 \tau ({\cal A}^{(1)(1)} + {\cal A}^{(2)(2)})}.
\label{concur}
\eeq
Taking into account Eqs.~(\ref{a11}) and~(\ref{a22}) it is easy to see that disorder contributes with a positive argument in the above exponential. As a consequence, the nonlocal decoherence can be strongly softened. In fact, assuming equal polarizations for the atoms, one notices that, for a fixed $\omega_0$, there is a critical value for the disorder strength $\sigma_{c}^{2}$ such that the initial entangled state behaves as a decoherence-free state:
\beq
\sigma_{c}^{2} = 12\pi/\omega_{0}^3.
\eeq
For $\sigma^{2} > \sigma_{c}^{2}$, ${\cal C}[\rho_{A}(\tau)]$ could be larger than one for $\tau > 0$ which is meaningless in the present case. Hence there must be a compromise between the atomic energy gap and the disorder strength so that the investigation of the entanglement by means of the concurrence yields physically acceptable results. A similar qualitative conclusion can be drawn for atoms with different polarizations.

It is possible to find an explicit expression for the concurrence, yet cumbersome, for a generic value of $\kappa$. For the sake of clarity we will not display such an expression here. In this case we will limit ourselves to a qualitative numerical description. The behavior of the concurrence as a function of the proper time $\tau$ of the atoms is depicted in Fig.~\ref{1} below. We choose the polarizations of the atoms such that
$$
\delta_{ij}\,p^{i}_{(1)}p^{j\,*}_{(1)} = \delta_{ij}\,p^{i}_{(2)}p^{j\,*}_{(2)} = 1.
$$
The conclusions are not qualitatively affected by considering an arbitrary configuration for the atomic polarizations. Observe the reduction in the entanglement decay as the intensity of disorder increases. Even though we do not observe a specific value for the disorder strength such that the concurrence remains constant for all $\tau$ as above, we do find a certain critical value for $\sigma^2$ from which the concurrence indicates creation of entanglement within a given proper time interval. Once again one must accept a sort of fine tuning between $\omega_0$ and $\sigma^2$ for a particular observational time interval so not to spoil the complete positivity of the time evolution. In the case of completely positive maps we are assured that the concurrence produces real values less than one. 
\begin{figure}[htb]
\begin{center}
\includegraphics[height=88mm,width=88mm]{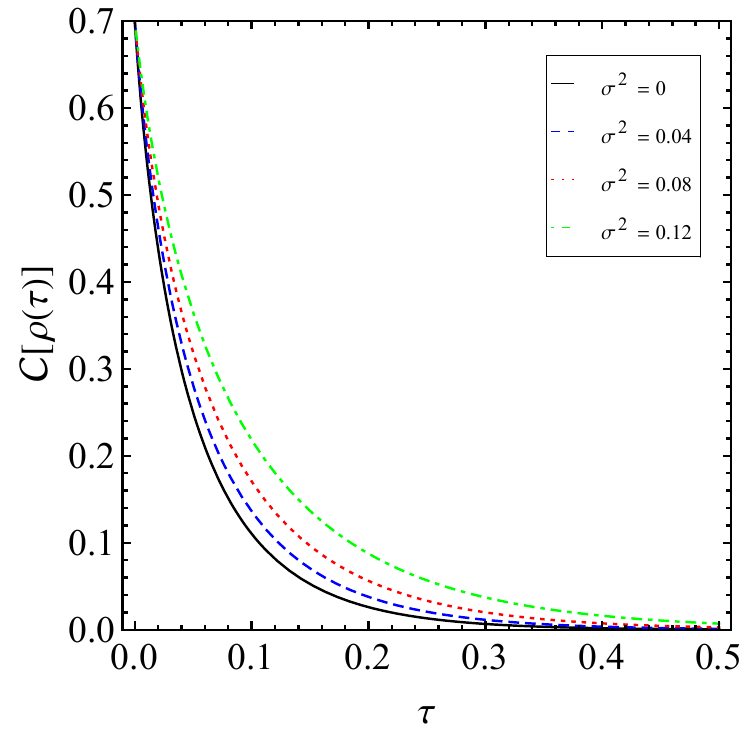}
\caption{The concurrence ${\cal C}[\rho(\tau)]$ as a function of $\tau$ for different values of the disorder strength $\sigma^2$. We choose $\kappa = 0.8$ and $\omega_0 = 5$. We remark that all physical quantities are given in terms of the natural units associated with the transition between the eigenstates of the atoms, with wavelength $\lambda = 2\pi/\omega_0$.}
\label{1}
\end{center}
\end{figure}

\vspace{-7mm}
\section{Conclusions}
\label{conclude}

In this paper we studied how the entanglement dynamics between two-level atoms is affected by random fluctuations of the light cone. In our model the atoms were coupled individually to two spatially separated cavities, one of which is called a disordered cavity. We demonstrated that in certain situations the disorder may contribute to enhance the degree of entanglement. This implies in a reduction in the entanglement decay and in turn that qubit systems can decohere slower in the presence of disorder. When the initial state of the qubits subsystem is given by the maximally entangled singlet state, we found a critical value for the disorder strength such that one observes an almost complete inhibition of the decay due to destructive interference. 

The physical picture that emerges in our context is the following. For atomic radiative processes, the spontaneous transition probability is reduced in the presence of random fluctuations of the light cone, which implies that atoms decay slower within a disordered environment. As a consequence, for entangled atoms, the interaction of one of such atoms with a noise field produces a destructive interference which hinders, to a certain extent, the transition to a separable state, resulting thereby in an enhancement of the mean life of the entangled state.

We believe that the approach investigated here for suppressing the effect of decoherence can be valid for other kinds of decoherence and for multipartite entangled states. In fact, we conjecture that in a strong disorder environment the mean life of entangled states can be enhanced in such a way as to nearly completely suppress nonlocal decoherence. Indeed, recently it was proposed that in a certain class of magnetic materials disorder generates quantum entanglement~\cite{savary}. Such outcomes corroborate our findings. 

A possible generalization of the present work is to investigate whether the same results are attainable in a non-perturbative framework. In turn, an alternative route to investigate the relationship between radiative processes of two-level atoms and disorder models is provided by the use of ultra-cold atomic systems. Indeed, from a soluble model of quantum optics, a spin-glass behavior can be obtained~\cite{Sarang, Sachev, Rotondo1, Rotondo2}. This spin-boson model describes a phase transition from the fluorescent to superradiant phase~\cite{andreev}. On the other hand, a recent mathematical rigorous method for the evaluation of the average free energy of a disordered system was developed in Ref.~\cite{nb}. The basic quantity is a distributional zeta-function. Within such a framework it has been verified the presence of a spontaneous symmetry breaking mechanism in a disordered $d$-dimensional Euclidean $\lambda\varphi^{4}$ model~\cite{prdisorder}. In addition, in the low temperature regime the existence of $N$ instantons in such a model has been confirmed. It has been discussed in the literature the possibility of representing a system of $N$ real instantons as $N$ macroscopic structures by $N$ two-level systems~\cite{Leggett}. The distributional zeta-function method could be an appropriate approach in order to demonstrate the emergence of an effective model of a bosonic field interacting with a reservoir of $N$ identical two-level systems within a disordered model.  Such subjects are under investigation by the authors.

\vspace{-4mm}
\section*{Acknowledgements}

This paper was partially supported by Conselho Nacional de Desenvolvimento Cient\'ifico e Tecnol{\'o}gico (CNPq, Brazil).

\end{document}